# A generative adversarial approach to facilitate archival-quality histopathologic diagnoses from frozen tissue sections

Kianoush Falahkheirkhah[1,2], Tao Guo[7], Michael Hwang[9], Pheroze Tamboli[7], Christopher G Wood[10], Jose A Karam[8,10], Kanishka Sircar[7,8], Rohit Bhargava[*,1,2,3,4,5,6]

[1] Department of Chemical and Biomolecular Engineering, University of Illinois at Urbana- Champaign, Urbana, IL 61801
[2] Beckman Institute for Advanced Science and Technology, University of Illinois at Urbana- Champaign, Urbana, IL 61801
[3] Department of Bioengineering, University of Illinois at Urbana- Champaign, Urbana, IL 61801
[4] Department of Electrical and Computer Engineering, University of Illinois at Urbana- Champaign, Urbana, IL 61801
[5] Mechanical Science and Engineering, University of Illinois at Urbana- Champaign, Urbana, IL 61801
[6] Cancer Center at Illinois, University of Illinois at Urbana- Champaign, Urbana, IL 61801
[7] Department of Pathology, University of Texas MD Anderson Cancer Center, Houston, TX
[8] Department of Translational Molecular Pathology, University of Texas MD Anderson Cancer Center, Houston, TX
[9] Department of Pathology and Laboratory Medicine, Indiana University School of Medicine, Indianapolis, IN
[10] Department of Urology, University of Texas MD Anderson Cancer Center, Houston, TX
[*]rxb@illinois.edu

# Abstract

In clinical diagnostics and research involving histopathology, formalin fixed paraffin embedded (FFPE) tissue is almost universally favored for its superb image quality. However, tissue processing time (>24 hours) can slow decision-making. In contrast, fresh frozen (FF) processing (<1 hour) can yield rapid information but diagnostic accuracy is suboptimal due to lack of clearing, morphologic deformation and more frequent artifacts. Here, we bridge this gap using artificial intelligence. We synthesize FFPE-like images ("virtual FFPE") from FF images using a generative adversarial network (GAN) from 98 paired kidney samples derived from 40 patients. Five board-certified pathologists evaluated the results in a blinded test. Image quality of the virtual FFPE data was assessed to be high and showed a close resemblance to real FFPE images. Clinical assessments of disease on the virtual FFPE images showed a higher inter-observer agreement compared to FF images. The nearly instantaneously generated virtual FFPE images can not only reduce time to information but can facilitate more precise diagnosis from routine FF images without extraneous costs and effort.

# Introduction

Histopathologic examination of microscopic morphologic patterns within stained tissues is often a critical diagnostic step in many clinical and research activities. Notably, the content and quality of images is dependent on the selection two methods to prepare thin sections - formalin-fixed, paraffin-embedding (FFPE) or fresh frozen (FF). FFPE processing is commonly used for stability of tissue to make it broadly available, requires fewer resources, enables easier sample preparation and better preserves tissue morphology but is time-consuming. Pathologists are historically accustomed to examining stained FFPE sections and determinations made on them are the typical gold standard. In contrast, FF processing can enable rapid decision making at the point of care and is more suitable for molecular analyses[1,2] but needs greater resources and may suffer from artifacts. FF techniques are especially suitable for intraoperative pathology since images can be obtained significantly faster (<30 minutes) compared to FFPE approaches (>24 hr)[3,4]. While FS processing allows decision-making in near real-time,[5] diagnoses are more challenging due to greater prevalence of artifacts and variability compared to archival quality FFPE images. This gap in diagnostic quality may lead to a deferral or inaccurate diagnosis.[6–11] In the case of kidney cancer, for example, FF cause difficulties in diagnoses[12] and discordance with FFPE gold standards[13], leading to an uncertain benefit for intraoperative assessment[14]. Here, we report bridging this gap by utilizing artificial intelligence (AI) to model the underlying relationship between morphology and contrast of FFPE and FF stained images.

Deep learning (DL)[15] is a powerful AI technique with successes in a variety of fields, including image processing[16], speech recognition[17], self-driving cars[18], and healthcare[19]. Instead of relying on extensive information to develop our DL framework, as is common for unconstrained problems, we sought to take advantage of the dual preparation methods in both the choice of the DL framework and our study design. Generative adversarial networks (GANs)[20] are a special class of DL methods that have garnered attention in image style transfer and synthesis[21–24]. They have also shown great promise for a variety of digital pathology tasks such as tumor classification[25–27], stain normalization[28–30], and virtual staining of label-free tissues[31–33]. We chose GANs to generate "virtual FFPE" images from FF images since they can be especially powerful in our study design. We use renal cell carcinoma (RCC) as an exemplar of the general problem in pathology of relating FS to FFPE since pathologists are familiar with evaluating both types of images and extensive ground truth is available. Moreover, RCC is not as prevalent as some of the major cancers but provides a large enough cohort to develop and validate the approach here. Since a paired FF-FFPE dataset from the same histological section is not feasible, we employed the next best available option - FF and FFPE processed sections from the same kidney, as close as possible to each other. Appropriate to this constraint, we developed a Cycle-GAN[30] framework for unpaired image-to-image translation. A typical GAN constructs two different convolutional neural networks: a generator and a discriminator. While the discriminator tries to distinguish real images from fake ones, the generator learns to produce images that are difficult for the discriminator to distinguish. Cycle-GANs, in contrast, employ generators and discriminators for each of the two domains so that the images can be translated between them. Here, we first modify the architecture of the generator and discriminator to rapidly convert FF to FFPE images and synthesize important finer morphology. The generator is a combination of U-Net[34] and Res-Net[35] and has been successfully implemented previously[36]. The discriminator is multi-scale, which is suitable to recover finer details[23]. Second, we evaluate the AI results using evaluations by board-certified pathologists who routinely examine FF and FFPE images from the same patients and provide the best human judges to relate results for realism and utility.

## Methods

### 1. Sample preparation

Archived formalin-fixed, paraffin-embedded tissues and frozen samples from The University of Texas MD Anderson Cancer Center (Houston, TX) were obtained after informed consent and using an institutional review board-approved protocol (IRB# LAB 08–670). For this retrospective study, we searched our institution's database for surgically resected kidney cases using the keywords "partial nephrectomy", "clear cell renal cell carcinoma", between 2019-2020. We found 40 cases of clear cell renal cell carcinomas that comprised the study cohort. Lesional tumor foci as well as non-neoplastic kidney controls were marked on hematoxylin- and eosin-stained slides from all cases. Briefly, 98 samples were extracted from 40 patients comprising 49 frozen sections and 49 adjacent permanent sections (FFPE) from the same tissue block. The patients age-range is from 22 to 81, including 13 females and 27 males.

## 2. Model design

The overall objective of the proposed study is to learn the translation between FF and FFPE domains. We introduce our framework in Fig. 1. After digitizing the FF and FFPE samples, we extract patches at 20x. Our framework follows Cycle-GAN[30], which includes two generators and two discriminators ($G_{FFPE}$, $G_{FF}$, $D_{FFPE}$, and $D_{FF}$) for translation between FF and FFPE domains. The $G_{FFPE}$ is responsible for translating images from FF to FFPE domain whereas $G_{FF}$ reverses the process and maps images from the FFPE domain to FF. Simultaneously, the discriminators send feedback on whether the images are real or fake. $D_{FF}$ takes FF domain images and $D_{FFPE}$ takes FFPE images. The proposed framework needs to be trained using proper loss functions for achieving stable and decent performance. For calculating the GAN loss, we use 2 discriminators that have an identical layout and we apply this to different image scale similar to the previous study[23]. One of them operates on a full resolution image while the other operates on images down-sampled by a factor of 2. These multi-scale discriminators have been rarely used in biomedical imaging. However, they can significantly improve the analysis of histological images as they evaluate the images at different field of views and resolutions -- similar to how pathologists make decisions. Moreover, we utilize a modified U-Net architecture that has been successfully used in the previous study[37]. These are modifications to the original Cycle-GAN framework, which improves the quality and better transfers the morphology.

## 3. Loss functions

We optimized the parameters of our framework with respect to a combination of two different loss functions: cycle loss and adversarial loss. The main idea behind cycle-GAN is to translate between domains through a cycle. In our case, the cycle is FF → Fake FFPE → Reconstructed FF and simultaneously, FFPE → Fake FF → Reconstructed FFPE. It is necessary to minimize the difference between reconstructed images and original images, which is called cycle loss and defined as:

$$\mathcal{L}_{cycle}\{G_{FF}, G_{FFPE}\} = ||G_{FF}(G_{FFPE}(FF)) - FF||_1 + ||G_{FFPE}(G_{FF}(FFPE)) - FFPE||_1 \quad (1)$$

For the translation from FF to FFPE, the adversarial loss is defined as:

$$\mathcal{L}_{adv}\{G_{FFPE}\} = (1 - D_{FFPE}(G_{FFPE}(FF)))^2 \quad (2)$$

$$\mathcal{L}_{adv}\{D_{FFPE}\} = (1 - D_{FFPE}(FFPE))^2 + D_{FFPE}(G_{FFPE}(FF))^2 \quad (3)$$

Where equations 2 and 3 represent adversarial loss for generator and discriminator, respectively. Similarly, for mapping from FFPE to FF, $\mathcal{L}_{adv}\{G_{FF}\}$ and $\mathcal{L}_{adv}\{D_{FF}\}$ are calculated in the same way. The overall objective function is achieved by linear combination of aforementioned loss as below:

$$\mathcal{L}_{total}\{G_{FF}, G_{FFPE}, D_{FF}, D_{FFPE}\} = \mathcal{L}_{adv}\{D_{FF}, D_{FFPE}, G_{FF}, G_{FFPE}\} + \gamma_1 \mathcal{L}_{cycle}\{G_{FF}, G_{FFPE}\} \quad (5)$$

Where $\gamma_1$ is regularization terms used to incorporate the importance of different losses to the total objective function and is set to 10 similar to original cycle-GAN[21].

## 4. Implementation details

The framework is implemented in PyTorch 1.3, CUDA 10.1, and Python 3.7.1 and computations are performed on a single NVIDIA® GeForce® RTX 2080 SUPER GPU and Intel(R) Xeon(R) Silver 4216 CPU @ 2.10GHz. All of the generative and discriminative models are initialized to random weights. Adam[38] was used to optimize the parameters of models with an initial learning rate of $10^{-4}$ and they are multiplied by 0.96 after every 1000 iterations. The training stops after a total 100,000 iterations. For data augmentation, we apply random affine transformation to the images at each iteration. The implementation can be found here: https://github.com/kiakh93/Virtual_FFPE

# 5. Evaluation

The evaluation of results is not straightforward since FF and FFPE are from different tissue sections and there is no one to one comparison between virtual FFPE and real FFPE. Thus, pixel level evaluation methods such as PSNR and RMSE cannot be used. Instead, we use two different approaches to evaluate our results: pathologist review and perceptual similarity.

## 5.1 Pathologist review

We created a survey to identify the quality of diagnoses. The purpose of this survey was to evaluate the quality and diagnostic reliability of virtual FFPE images compared to real FFPE. The survey included 35 FF, 35 virtual FFPE, and 35 real FFPE samples. The virtual FFPE images have been generated given the FF images using the generative network ($G_{FFPE}$), whereas the real FFPE images have been selected from parallel sections corresponding to the FF samples. Each image in the survey has field of view of 2mm x 2mm. Board-certified pathologists were asked to review each of the images and to answer two questions:

- How closely does the image resemble FFPE tissue, on a scale of 1to 5? (1 is a FF section and 5 is a permanent FFPE section)
- What is the grade of cancer?

The first question quantitatively evaluates the similarity between virtual FFPE images and real FFPE images. Hence, for each group of images (virtual FFPE, FFPE, FF) we will have an average score between 1 and 5 which indicates the similarity of each group to FFPE or FF domains. The second question, however, measures pathologists' concordance rate in grading clear cell renal carcinoma for FFPE, virtual FFPE, and FF samples. The hypothesis is that the virtual FFPE images increase the concordance rate compare to FF images, which we can be tested by comparing their Fleiss' kappa[39]. This measure quantifies the degree of agreement in classification above that which would be expected by chance. A value of 0 demonstrates agreement occurring only by chance and a value of 1 shows perfect agreement. Upon completion of the survey by pathologists, we performed statistical analyses for both the concordance rate of diagnostic interpretation and inter-observer variability (question 2 in the survey), we use Fleiss' kappa statistic, which is a common statistical approach for analyzing inter-observer variability.

## 5.2 Perceptual similarity

There has been an ever-present challenge to perceptually compare quality of images. The most common objective perceptual metrics, such as peak signal-to-noise ratio (PSNR) and structural similarity index measure (SSIM)[40], fail to consider many perceptual differences detected by humans. A new metric has recently been introduced[40] as an alternative for DL to systematically compare deep features of images and is called Learned Perceptual Image Patch Similarity (LPIPS) metric. This metric seeks to quantify how perceptually apart two images are; a higher LPIPS value indicates a greater dissimilarity. We calculated the LPIPS metric for 2000 image patches each for FF, the corresponding virtual FFPE, and FFPE. The size of image patches was 512 x 512. In our case, the reference images were real FFPE and we calculate LPIPS for FF and virtual FFPE images.

# Results

Since FF and FFPE samples are from the same patient but not serial sections, they are not exactly matched; we developed a cycle-GAN[21] framework for this unpaired collection (Fig. 1). To train our framework, we used a dataset that includes hematoxylin and eosin (H&E)-stained whole slide images of FF sections and FFPE sections from the same 20 samples (with a total of 6396 patches of 512 x 512) across different grades of clear cell RCC. The model estimates FFPE images after training from FF images alone, without the need for other FFPE images or any further information. The key elements of these virtual FFPE images involve morphological structures, staining levels and contrast, each of these can be appreciated in Fig. 2 for a benign and a cancerous sample. These results demonstrate that the model is capable of translating FF sections (Fig. 2A) into virtual FFPE (Fig. 2B), showing the structural and morphological details that may be expected from real FFPE images (Fig. 2C). Additional examples can be found in supplementary Figure 1. The inference itself is fast, requiring 0.105 seconds for a 2000 x 2000 pixel region, which can be further improved by using stronger computing hardware with parallelization capabilities. We especially note that virtual FFPE images can be expected to have a more consistent appearance over a population, uniformity in color and contrast across the image and a tunable level that will likely not be matched by real stains. Although visual agreement is good between virtual and real FFPE images in terms of morphologic patterns, color and contrast, there are also invariable differences between the precise structures in these images as they are from different parts of the tissue. A quantitative evaluation is needed to assess the quality and utility of the predictions.

We used determinations of practicing pathologists as an independent test of the results. Five pathologists completed the survey, as described in the methods section, to assess images. The pathologists are board certified, practicing clinicians with experience: KS (25 years); TG (5 years); MH (6 years); AA (5 years); BN (6 years). The survey was designed with 35 examples each of FF, virtual FFPE, and FFPE samples to ensure that a large number of images were analyzed but not so large as to induce fatigue that may introduce error during the tests. The set was the same for all pathologists but was randomized in the order of presented images presented to each pathologist. In order to quantify the assessment on a common scale, we developed a Pathologist Evaluation Index (PEI) that assigns a 1-5 score to each image. Here, 1 indicates how closely the presented field of view from the image resembled an FF image and 5 an FFPE sample. As summarized in Table 1, pathologists assigned lower and higher PEI scores to FF and FFPE images, respectively, serving to assure the relative magnitudes of the assessment and bounding the expected range for both the test and individuals. PEI scores higher than 1 for FF and lower than 5 for FFPE likely arise from the limited fields of view that we had to present to each pathologist to ensure that the regions they were looking at were consistent. The intermediate scores may also reflect that each pathologist has knowledge that these images are from the three sources. The distribution of scores helps bound the effects from this bias and individual scoring proclivities, emphasizing that the raw scores are not globally absolute. In all cases, importantly, we note that the virtual FFPE scores were intermediate between the FF and FFPE. Artifacts, for example from freezing and sample folding, persisted in the virtual FFPE images; consequently, their average score is lower than the one assigned to FFPE by pathologists on average. It is possible, in principle, to use AI methods to eliminate these known artifacts and their effects in virtual FFPE. While we show this possibility in the discussion section, we did not attempt to optimize diagnostic quality in this study to keep it focused on FF-FFPE translation.

While physical quality assessment is reassuring, the impact of this image on pathologic assessment is important. In particular, detecting tumor for margin assessment is a valuable application of FF processing. Hence, we focused next of evaluating the performance of pathologists in distinguishing benign versus cancerous regions by calculating inter-observer agreements using Fleiss' kappa. Further, to assess the utility in characterizing disease we also examined their determination of tumor grade for clear cell RCC (question 2 in the survey). As shown in Table 2, Fleiss' kappa value (κ) for FF, virtual FFPE, and FFPE images is 0.52, 0.67, and 0.89, respectively, for distinguishing benign versus cancerous samples. For grading clear cell RCC, it is 0.39, 0.51, and 0.63, respectively. A kappa value 0 demonstrates agreement by chance and 0.01–0.20 as none to slight, 0.21–0.40 as fair, 0.41–0.60 as moderate, 0.61–0.80 as substantial, and 0.81–1.00 as almost perfect agreement[41]. Thus, our findings suggest that pathologists' agreement for virtual FFPE images increases from moderate to substantial for benign versus cancerous and from fair to moderate for grading clear cell RCC. This is not surprising as discriminating benign from cancerous tissue is much easier than assigning grade, which shows greater interobserver variability[42,43].

Finally, we used automated and objective algorithms used in computerized image processing to calculate the perceptual similarity between the three classes of images using a pre-trained convolutional neural network as a perceptual similarity metric that mimics human judgment[40]. This Learned Perceptual Image Patch Similarity (LPIPS) was measured for 2000 image patches of FF versus FFPE and virtual FFPE versus real FFPE. Among those images, more than 93% of virtual FFPE patches were assigned to be more similar to the corresponding FFPE patches by LPIPS metric. Thus, as summarized in Table 3, virtual FFPE images have lower average LPIPS value compare to FF images indicating stronger similarity between virtual FFPE and FFPE images.

# Discussion

In this study, we present a DL framework that facilitates rapid and more assured diagnoses by combining the speed advantages of FF with the quality of FFPE processing. Without any additional tissue processing of FF sections, our computational framework transfers the morphologic textures and colors from FF samples to that similar to those obtained after FFPE processing in the form of virtual FFPE images. Overall, the use of virtual FFPE images increases inter-observer agreement between pathologists compared to using FF alone; importantly, without increasing workflow times, training or using any additional reagents. This is not unexpected as images that resemble the gold standard (FFPE) offer the potential for more precise diagnosis with fewer complicating factors. The presented approach, thus, offers a significant value for the small effort needed in implementing and using it. The present study involved examining a specific organ and tumor type; however, our strategy may be employed to generate images for other organs and diseases with appropriate DL framework tuning. Hence, the pathologists' workflow may be facilitated when evaluating FF samples for a variety of indications where faster, more assured information is desirable, including diagnosis and subtyping of tumors, assessment of surgical resection margin status, evaluation of allografts for organ transplantation and screening at the time of tissue procurement for ancillary testing.

To better transfer the morphology and remove artifacts, we modified architectures and loss functions of the original cycle-GAN[30]. We utilized a generator that is a combination of U-Net[34] and Res-Net[35] architectures. Our generator leverages advantages of both architectures such as learning multi-level features and preventing the vanishing gradient problem. Furthermore, we used 2 discriminators with identical architecture that operate at different image scales similar to previous studies[44]. One of the discriminators operates at a larger receptive field, guiding the generator to synthesize images with consistent global context. The other discriminator, however, performs locally, which encourages the generator to synthesize finer morphology and semantically valid structures that exists in FFPE images. Following the aforementioned modifications, we observed that virtual FFPE images look more realistic and better smooth out freezing artifacts (supplementary Figure 3) compared to the original cycle-GAN. In addition, the LPIPS value of these images (0.45 ± 0.07) further confirms the superior performance of our designed framework compared to the original cycle-GAN. A comprehensive examination and strategies for elimination of artifacts is possible and contributions of such efforts can now be evaluated in larger sample sets since we show the possibility of the primary FF-FFPE translation.

Similar to every other machine learning model, our framework has some potential failure cases. For instance, occasional artifacts from FF domain persist in virtual FFPE images, such as the sample being folded or torn (supplementary Figure 4). Obviously, a style transfer approach cannot repair physical damages but an additional module to our workflow that focuses on inpainting can potentially help[45]. There are other artifacts that are also well know. For example, when the freezing process is slow, cell morphology may not be well preserved and cells typically appear larger. The problem of bloated cell morphology cannot be fully resolved using our framework at present though further extensions can be made to address known artifacts. Finally, in order to apply this to other tissue types, training on the corresponding tissue datasets is necessary for accurate synthesis of virtual FFPE images. Different types of tissues include different cell types, morphologies, and structures, whose textures have to be introduced to the prediction framework at the training stage. In addition, different institutions use different protocols, sample preparations, and imaging scanners, which cause color variation in histological images. Thus, various type of tissues from different institutions are needed to generalize translation between FF and FFPE domains. We anticipate that a larger number of samples and greater diversity can potentially lead to better training of the model and also make the predictions more powerful.

Our method is very rapid; for example, for a 1 cm x 1 cm image with 0.5 μm pixel sizes, the computation takes approximately 10.5 seconds on a single GPU. Given that modern optical scanners can digitize a high-resolution image within minutes, our approach has the potential for real-time analysis during intraoperative consultation. As most intraoperative frozen sections involve interpretation of fewer than 5 slides per case, the extra time required using our approach would be a few minutes. In practice, a pathologist will use his/her judgement to decide if it is worthwhile to perform FF to virtual FFPE translation in a given case, but we do not anticipate that the extra few minutes will not be a barrier. Pathologists may use this technique for an organ site or tumor type that is known to be problematic intraoperatively, or a particularly challenging case, or if a more precise diagnosis is desired. Just as ordering a recut section or calling for a colleague's opinion, our method could prove to be a valuable tool in the arsenal of the pathologist. This method also suggests analysis strategies that may be powerful; for example, performing FF to FFPE translation on one slide while waiting for a recut section or other sections from the same case to be cut and stained. Ultimately, the conclusion is that this tool provides a means to allow the pathologist the option to evaluate a near FFPE quality section without waiting until the following day. In order to further evaluate the impact of virtual FFPE images on accuracy of diagnosis, a large-scale clinical evaluation is needed that includes a wide range of tissue types and pathologists with differing levels of expertise and experience to carefully analyze a large number of images to recognize other potential failure cases and drawbacks.

The technique presented here also presents new opportunities in addressing the diversity in pathologist capability for assessing frozen section pathology on two fronts. In general, the accuracy of diagnoses from FF sections improves with pathologist experience. Given that relatively few, mostly academic medical centers perform a high volume of FF sections, expertise is highly variable and can take long to acquire. An approach like the one presented can aid in development of expertise by allowing a user to analyze FF images and get its FFPE "twin". This approach can be especially useful in enabling diagnoses among the larger cohort of community pathologists[46], where opportunities to gain FF section pathology expertise are fewer as well as for organs in which disease is not especially prevalent. In contrast to highly prevalent cancers where opportunities for understanding and training can be significant, such as that of the breast, the use of our method for cancers with fewer occurrences can be more impactful. Another complementary aspect of our framework is generating virtual FF images given real FFPE images (supplementary Figure 2). While virtual FF images do not have additional diagnostic value for pathologists, they similarly offer significant value for educational purposes. New technologies to aid pathologist education can be useful where a concern may be quality of extensive training in a short period of time[47].

## Conclusions

We report a machine learning approach to bridge the gap between the two prevalent methods of tissue processing for histopathology. Using a deep learning framework, we report on this method to synthesize FFPE images of kidney samples, which we term virtual FFPE, from clinical FF images. Validation of the quality and utility of virtual FFPE images was assured using a survey administered to multiple board certified and experienced physicians, demonstrating that the virtual FFPE are of high quality and increase inter-observer agreement, as calculated by Fleiss' kappa, for detecting cancerous regions and assigning a grade to clear cell RCC within the sample. This framework can be broadly applied to other type of tissues across the biomedical sciences to generalize the mapping between FF and FFPE domains and derive the advantages of both. This study paves the way for routine FF assessment to be augmented with benefits of FFPE information without adding to the cost, time or effort required, thereby increasing the quality of histopathologic examinations using machine learning.

# Disclosures
There are no conflicts of interest to report.

# Acknowledgements
In addition to the co-authors, we would like to acknowledge pathologists who blindly completed this survey (AA ; BN). This work was supported by the National Institutes of Health via grant number R01CA260830.

|  | Assigning score by pathologists | | | | | |
| --- | --- | --- | --- | --- | --- | --- |
|  | 1 | 2 | 3 | 4 | 5 | Total |
| **Frozen** | 1.86 ± 0.94 | 3.26 ± 1.06 | 1.34 ± 0.87 | 1.46 ± 0.78 | 1.48 ± 0.77 | 1.88 ± 1.14 |
| **Virtual FFPE** | 2.53 ± 0.84 | 3.79 ± 0.47 | 2.53 ± 1.33 | 2.56 ± 0.93 | 1.87 ± 0.93 | 2.65 ± 1.18 |
| **FFPE** | 2.91 ± 1.01 | 3.94 ± 0.33 | 3.72 ± 1.25 | 3.28 ± 0.98 | 2.40 ± 1.14 | 3.25 ± 1.07 |

Table 1. Pathologists review of the survey for assigning score to the images, where we report mean ± standard deviation of the scores that have been assessed by pathologists to each domain. Scores interval is from 1 to 5, where 1 being FF-like samples and 5 being FFPE-like samples.

|  | Fleiss' Kappa with 95% CI | |
| --- | --- | --- |
|  | Benign vs cancer | Grading |
| **Frozen** | 0.52 (0.42,0.62) | 0.39 (0.34,0.44) |
| **Virtual FFPE** | 0.67 (0.59,0.75) | 0.51 (0.44,0.58) |
| **FFPE** | 0.89 (0.78,1.00) | 0.63 (0.56,0.70) |

Table 2. Pathologists review of the survey for calculation of inter-observer agreement for each domain using Fleiss' kappa for 95% confidence interval.

|  | LPIPS metric |
| --- | --- |
| **Frozen** | 0.47 ± 0.06 |
| **Virtual FFPE** | 0.41 ± 0.07 |

Table 3. Comparison of LPIPS metric for FF and virtual FFPE image patches. The lower LPIPS value the closer that domain is to the reference which is archival FFPE. We report mean ± standard deviation

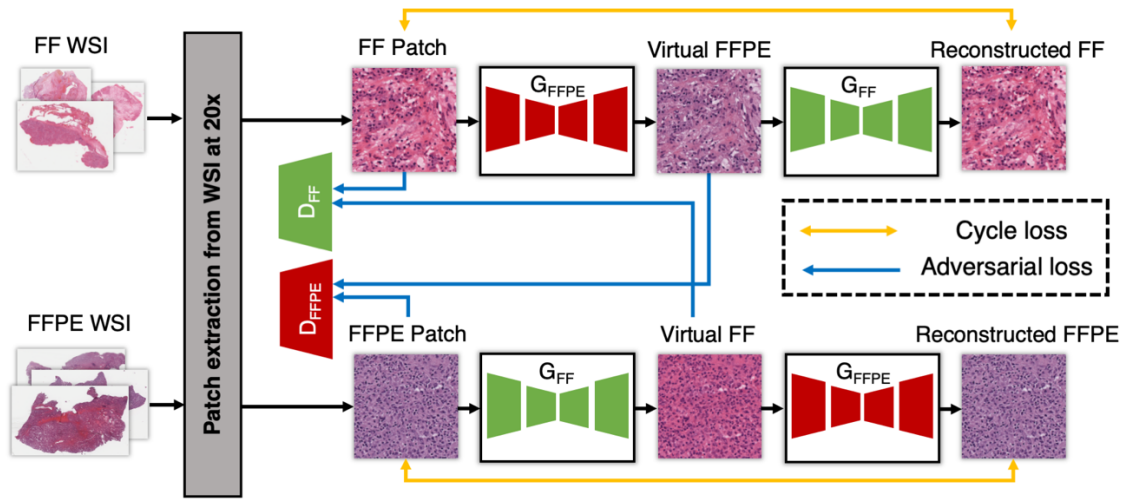

Figure 1. An overview of the proposed method. Our framework includes two generators and two discriminators for translation between FF and FFPE domains and vice versa.

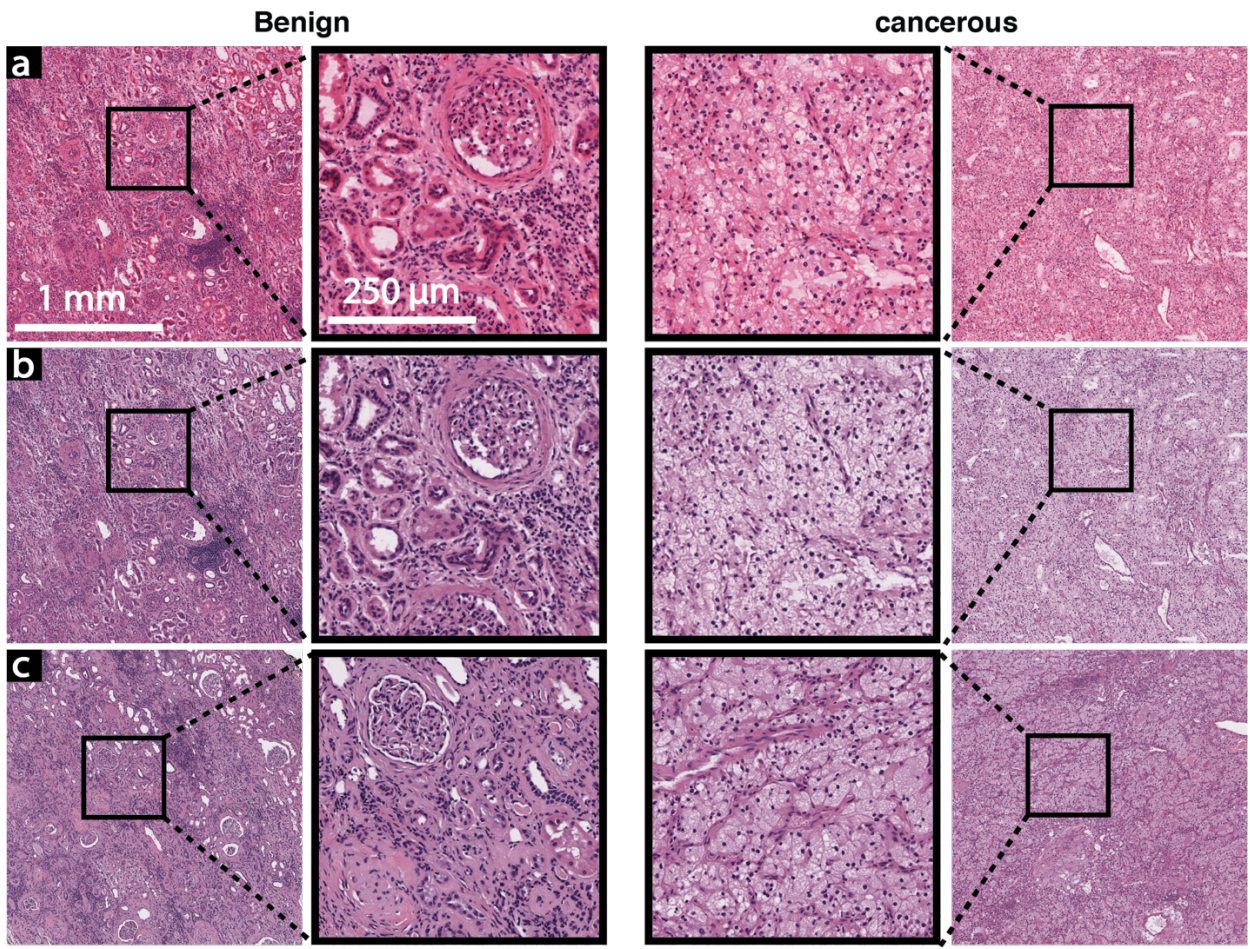

Figure 2. Visual quality assessment. a, FF images. b, Virtual FFPE images. c, Real FFPE images. From left to right we show benign kidney samples, at low and high magnifications and kidney cancer samples at low and high magnifications, respectively.

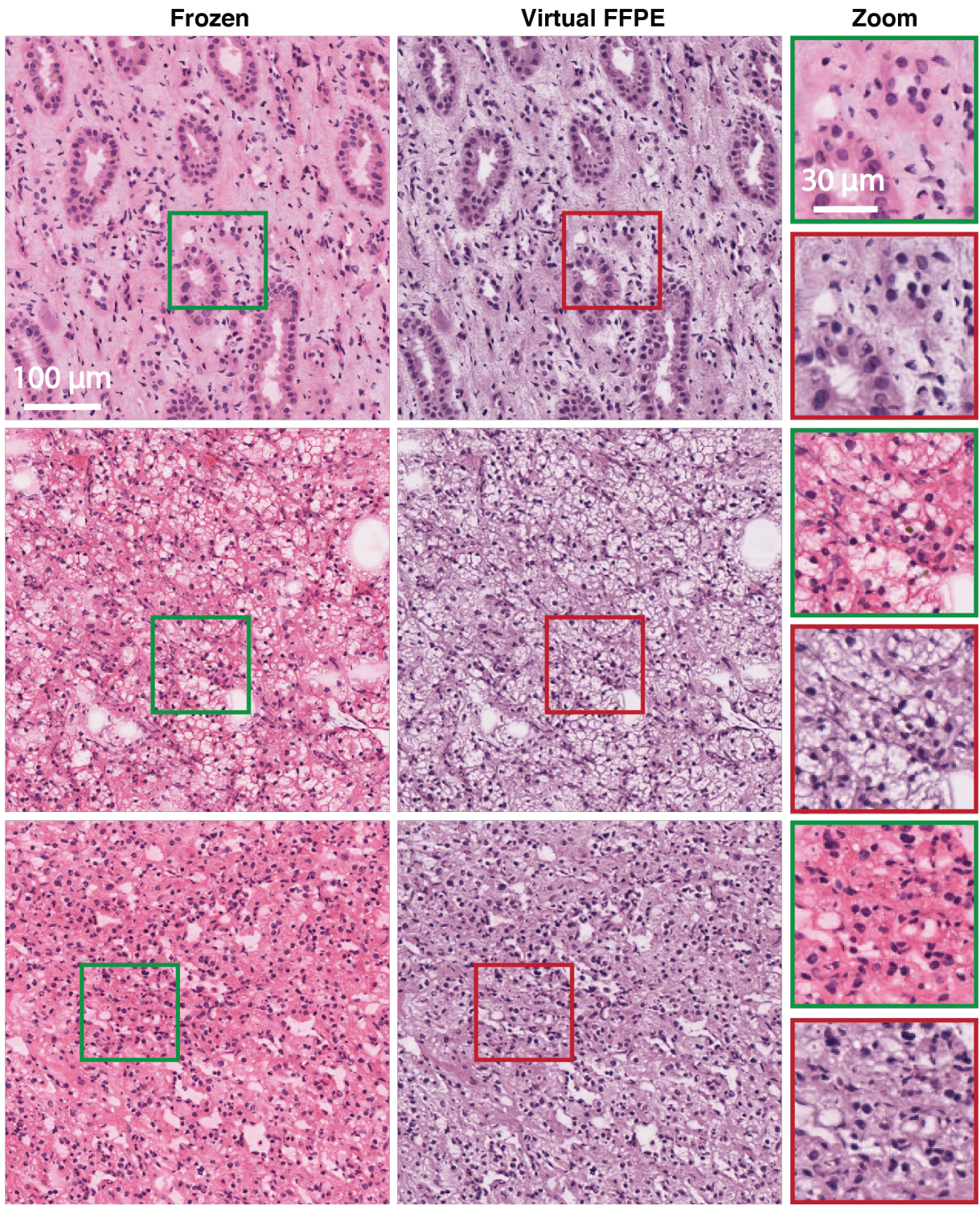

Supplementary Figure 1. More examples of the presented framework. The left column represents frozen samples that are input to the network. Them middle columns illustrates the virtual FFPE images. The right column shows the zoom area. Each row represents a different patient.

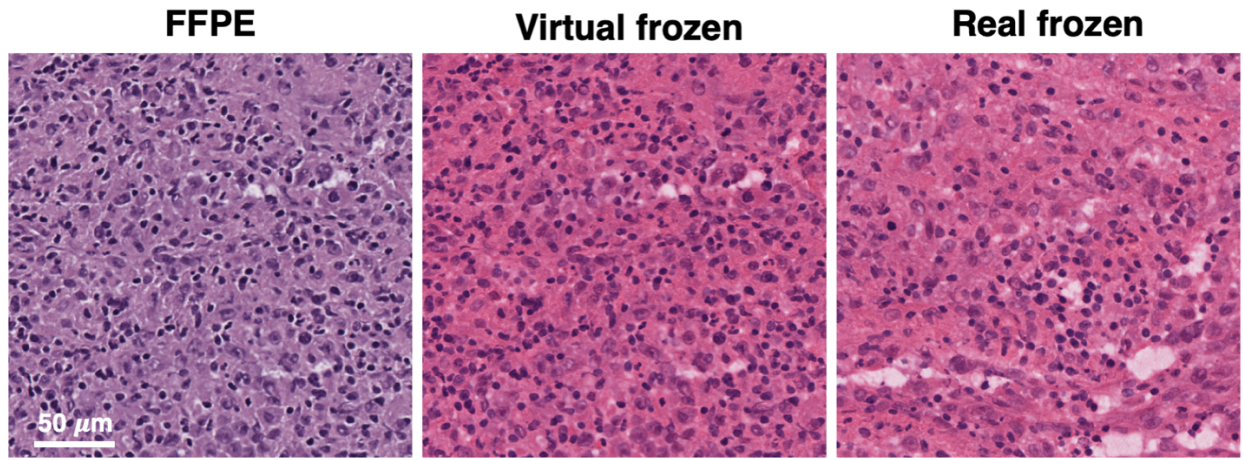

Supplementary Figure 2. An example of virtual FF image generated by $G_{FF}$. From left to right we show real FFPE image (input), virtual FF (output), and real FF image.

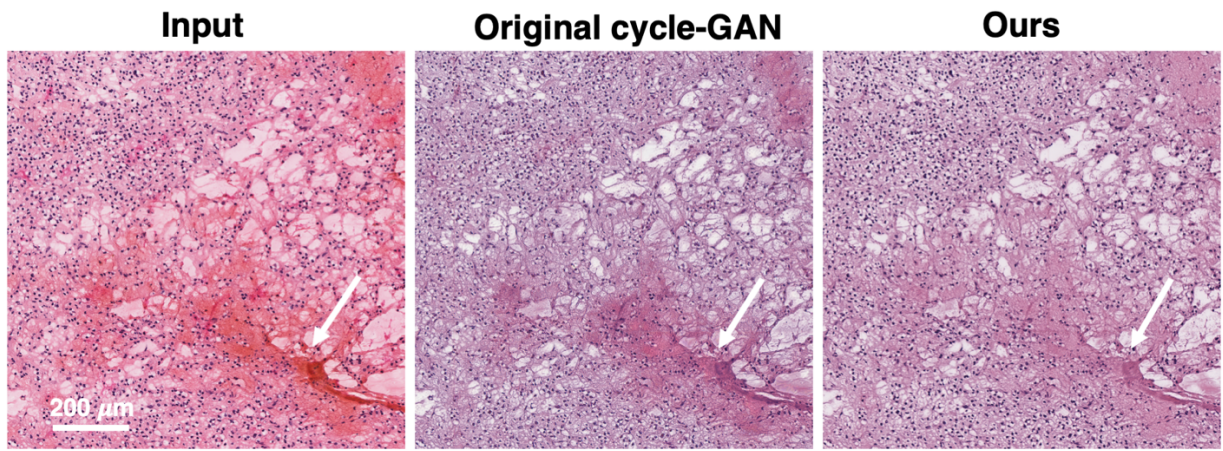

Supplementary Figure 3. Visual comparison of virtual FFPE images. From left to right we show the FF image as input, output of original cycle-GAN, and output of our framework. Our framework better smooths out freezing artifacts (arrows) and generates more realistic virtual FFPE images

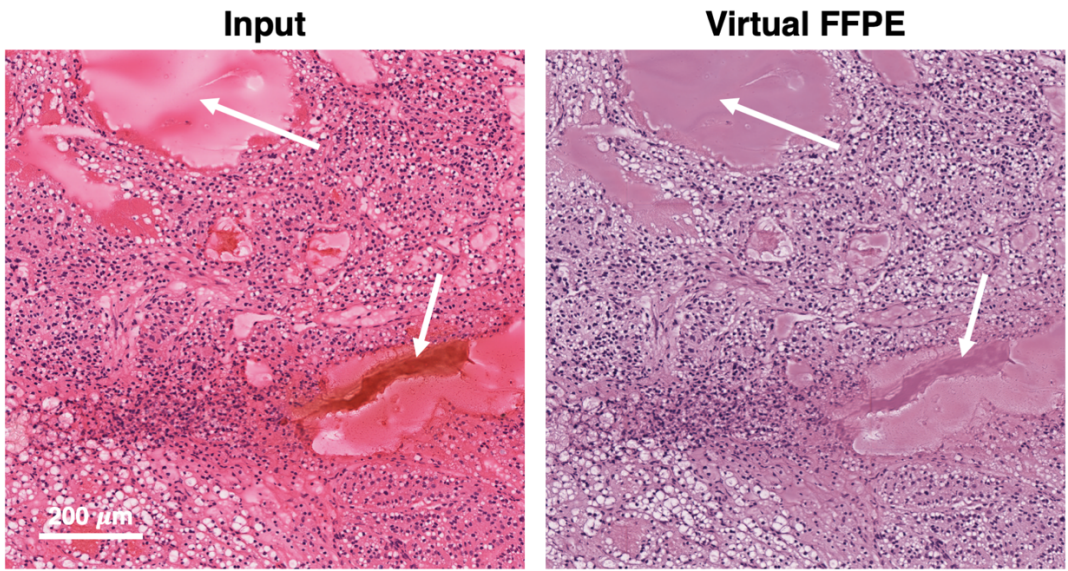

Supplementary Figure 4. Potential failure cases. Our framework cannot mitigate all of the freezing artifacts for all of the samples (highlighted with arrows).